\documentclass[aps,prb,twocolumn,superscriptaddress,showpacs]{revtex4-1}

\usepackage{ulem}
\usepackage[utf8]{inputenc}
\usepackage{graphicx}
\usepackage{dcolumn}
\usepackage{bm}
\usepackage{hyperref}
\usepackage{amsmath}
\usepackage{mathtools}
\usepackage[dvipsnames]{xcolor}
\usepackage{amssymb}
\usepackage{amsmath}

\begin{document}

\preprint{APS/123-QED}

\title{Anti-hyperuniform diluted vortex matter induced by correlated disorder}

\author{Joaqu\'{i}n Puig}
\affiliation{Low Temperatures Lab, Centro At\'{o}mico Bariloche, CNEA, Argentina.}
\affiliation{Instituto de Nanociencia y Nanotecnología, CONICET-CNEA, Nodo Bariloche, Argentina.}
\affiliation{Instituto Balseiro,
CNEA and Universidad Nacional de Cuyo, Bariloche, Argentina}

\author{Jazm\'{i}n Arag\'{o}n S\'{a}nchez}%
\affiliation{Low Temperatures Lab, Centro At\'{o}mico Bariloche, CNEA, Argentina.}
\affiliation{Instituto de Nanociencia y Nanotecnología, CONICET-CNEA, Nodo Bariloche, Argentina.}
\affiliation{Instituto Balseiro,
CNEA and Universidad Nacional de Cuyo, Bariloche, Argentina}

\author{Edwin Herrera}
\affiliation{Laboratorio de Bajas Temperaturas y Altos Campos Magnéticos, Departamento de Física de la Materia Condensada, Condensed Matter Physics Center (IFIMAC) and Instituto Nicolás Cabrera, Universidad Autónoma de Madrid,  Spain}

\author{Isabel Guillamón}
\affiliation{Laboratorio de Bajas Temperaturas y Altos Campos Magnéticos, Departamento de Física de la Materia Condensada, Condensed Matter Physics Center (IFIMAC) and Instituto Nicolás Cabrera, Universidad Autónoma de Madrid,  Spain}

\author{Zuzana Pribulov\'{a}}
\affiliation{Institute of Experimental Physics, SAS, Kosice, Slovakia}

\author{Josef Ka\v{c}mar\v{c}\'{i}k}
\affiliation{Institute of Experimental Physics, SAS, Kosice, Slovakia}

\author{Hermann Suderow}
\affiliation{Laboratorio de Bajas Temperaturas y Altos Campos Magnéticos, Departamento de Física de la Materia Condensada, Condensed Matter Physics Center (IFIMAC) and Instituto Nicolás Cabrera, Universidad Autónoma de Madrid,  Spain}

\author{Alejandro Benedykt Kolton}
\affiliation{Instituto Balseiro,
CNEA and Universidad Nacional de Cuyo, Bariloche, Argentina}
\affiliation{Condensed Matter Theory Group, Centro At\'{o}mico Bariloche, CNEA, Argentina.}

\author{Yanina Fasano}
\affiliation{Low Temperatures Lab, Centro At\'{o}mico Bariloche, CNEA, Argentina.}
\affiliation{Instituto de Nanociencia y Nanotecnología, CONICET-CNEA, Nodo Bariloche, Argentina.}
\affiliation{Instituto Balseiro,
CNEA and Universidad Nacional de Cuyo, Bariloche, Argentina}
\affiliation{Leibniz Institute for Solid State and Materials Research,  Dresden, Germany}

\date{\today}

\begin{abstract}
Disordered hyperuniform materials are very promising for applications but the successful route for synthesizing them requires to understand the interactions induced by the host media that can switch off this hidden order.
With this aim we study the model system of vortices in the $\beta$-Bi$_2$Pd superconductor where correlated defects seem to play a determinant role for the nucleation of a gel vortex phase at low densities. We directly image vortices in extended fields-of-view and show that the disordered vortex structure in this material is anti-hyperuniform, contrasting with the case of vortex structures nucleated in samples with point-like disorder. Based on numerical simulations, we show that this anti-hyperuniform structure arises both, from the interaction of a diluted vortex structure with a fourfold-symmetric correlated disorder quite likely generated when cleaving  the samples and from the out of equilibrium nature of the quenched  configuration.

\end{abstract}

\maketitle

\section{Introduction}

Disordered  patterns are ubiquitous in material and biological systems composed of interacting objects and come in two flavors with distinctive physical properties, each of them having a different behavior of density fluctuations with distance. Some  disordered systems are entailed with a hidden order named hyperuniformity~\cite{Torquato2003} or superhomogeneity,~\cite{Gabrielli2002} characterized by vanishing density fluctuations at large length-scales. Disordered hyperuniform structures are incompressible at thermal equilibrium and entailed with physical properties of practical interest such as being transparent in a wide range of wavelengths.~\cite{Torquato2018}
In contrast,  non-hyperuniform disordered systems do not present this hidden order since their large-wavelength density fluctuations do not vanish.
Non-hyperuniform structures are typically opaque to light and are compressible systems.~\cite{Torquato2018} Interestingly, some systems can pass from hyperuniform to non-hyperuniform and thus switch on or off interesting physical properties just by changing the type of quenched disorder of the host media where the interacting objects are nucleated.~\cite{Rumi2019,Puig2022,Puig2023} For instance, this switch off effect can be anisotropically induced in vortex matter when the host media presents parallel planes of disorder instead of point-like disorder.~\cite{Puig2022,Puig2023}
Thus, unveiling which interactions induced by the host media can switch off the hidden hyperuniform order of a system is very important for both, practical and basic reasons.

For hyperuniform systems the variance of the number $N$ of interacting objects grows with
the observation length-scale $r$
as $\sigma_{N}^{2} \propto r^{d - \alpha}$, with $d$ the dimension of the system and $\alpha >0$.~\cite{Torquato2003} This  implies that their structure factor   $S(\mathbf{q}) \propto q^{\alpha} \rightarrow 0$ in the asymptotic $\mathbf{q} \rightarrow 0$ limit.~\cite{Torquato2003} A non-exhaustive list of disordered hyperuniform systems includes the distribution of the density
fluctuations in the early Universe,~\cite{Gabrielli2003} biological
tissues,~\cite{Zheng2020c} patterns of
photoreceptors in avian retinas,~\cite{Jiao2014} bubbles in foam,~\cite{Chieco2021}
 jammed particles,~\cite{Zachary2011,Dreyfus2015} two-dimensional material
structures,~\cite{Man2013,Chen2018,Zheng2020,Salvalaglio2020,Chen2021,Chen2021b}
and vortex matter in type-II
superconductors.~\cite{Rumi2019,Llorens2020b,Puig2022,AragonSanchez2022,Besana2024}
In contrast, many other disordered systems do not present this hidden order,
like normal fluids that have $\sigma_{N}^{2}\propto r^{d}$ and the $S(q\to 0)$ is bounded. Among the non-hyperuniform systems there are some where $\sigma_{N}^{2}$ grows faster than $r^d$ and $\alpha<0$, the so-called anti-hyperuniform structures.~\cite{Torquato2021}
Anti-hyperuniform matter includes fractals, systems at critical points,~\cite{Nishimori2011}  one-dimensional substitution tilings,~\cite{Oguz2019} and metastable states in densely-packed spheres.~\cite{Frusawa2021}

\begin{figure*}[ttt]
\centering
\includegraphics[width=2\columnwidth]{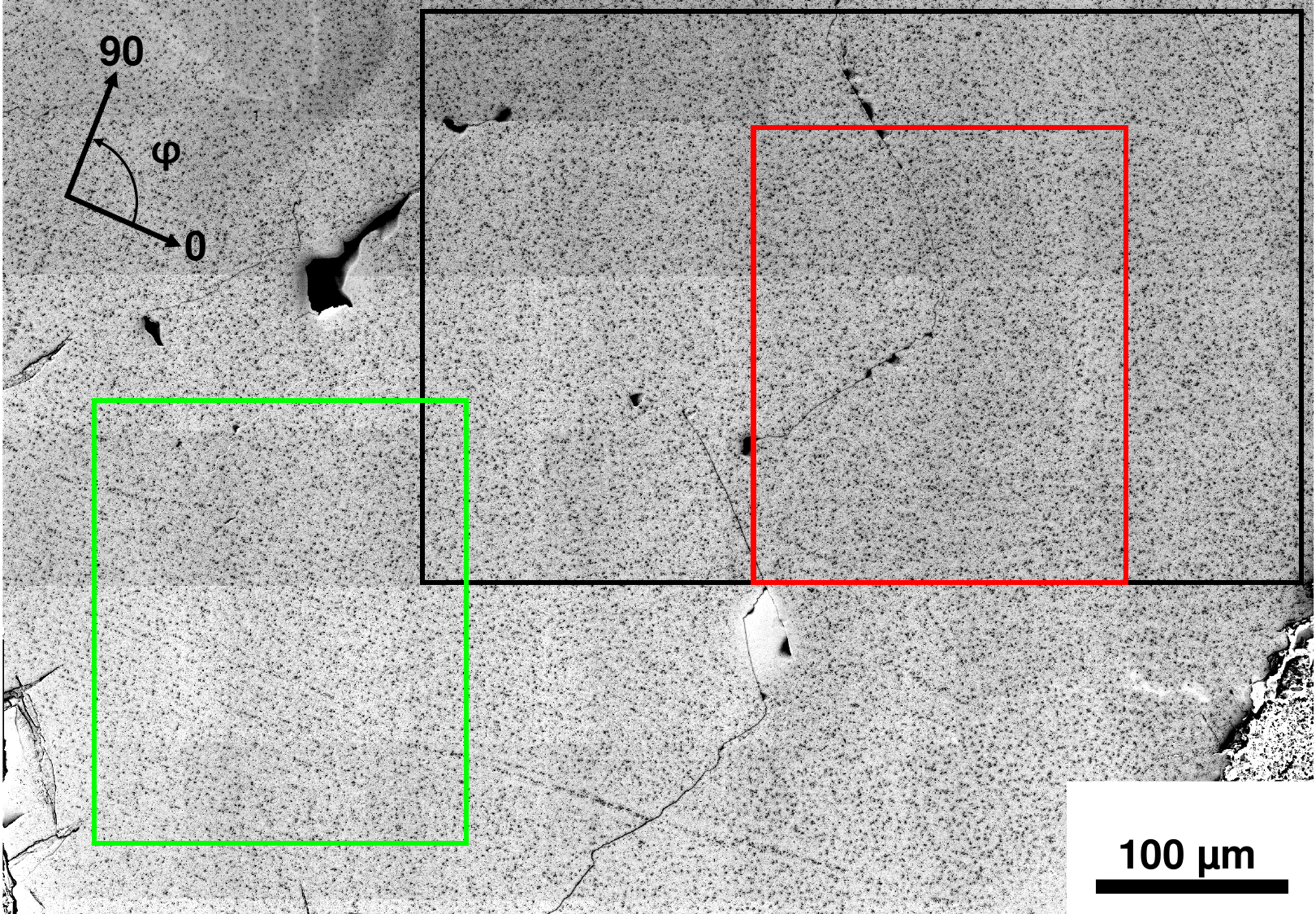}
\caption{Panoramic of the vortex structure  nucleated at low densities of 0.9\,G in a cleaved $\beta$-Bi$_2$Pd sample. The field-of-view spans  17 000 vortices (black dots) individually imaged in a magnetic decoration experiment performed at 2.3\,K after following a field-cooling process. The green frame highlights a region of the sample where perpendicular rows of vortices are clearly observed along the $\varphi =0$ and 90 degrees directions. The red frame highlights a region where these perpendicular vortex rows are less evident. The black framed region indicates the largest rectangular region of the sample considered in our analysis, spanning a field-of-view of 9 000 vortices. }
\label{fig:Figure1}
\end{figure*}

Here we use vortex matter in superconductors as a model elastic system to
investigate which interactions can eventually transform a system into a non-hyperuniform one. Vortex matter is hyperuniform if nucleated in media with point-like disorder.~\cite{Rumi2019,Llorens2020b,AragonSanchez2022}  Vortex-vortex interactions in most  three-dimensional superconductors are well described by  the London short-ranged potential,~\cite{Blatter1994,Demirdis2011,vanderBeek2012,Bolecek2016,AragonSanchez2020} but in some tetragonal materials the magneto-elastic effect is strong and the vortex-vortex interaction has an extra long-ranged term.~\cite{Lin2017} In the latter case the hyperuniformity class of the system is weakened even though disorder is point-like.~\cite{AragonSanchez2022} More extremely, the hyperuniform properties of vortex matter are anisotropically depleted if the sample has planar disorder.~\cite{Puig2022}
In addition, quenched disorder in the sample leads to glassy properties at low temperatures, such as slow relaxation and deep metastable states, whose effects are non-negligible in the density fluctuations at large length-scales.~\cite{AragonSanchez2022,Puig2022} Then vortex matter is a playground to investigate  which interactions might switch off the hidden hyperuniform order so interesting for applications.

Our main motivation here is to explore the possibility of switching off the hyperuniform hidden order observed in some materials by changing substantially the nature of disorder and the density of constituents of the system. Remarkably,  we have found that hyperuniformity can not only be switched off but also systems with anti-hyperuniform properties can be nucleated at diluted densities in host media with a particular correlated disorder. Understanding the mechanism of nucleation of anti-hyperuniformity is also another motivation of this work. Thus we study the effect of both, the disorder present in the host media and the freezing dynamics in the possible switching off of hyperuniformity.
The main result of this work is that a fourfold-symmetric correlated disorder potential can turn the otherwise hyperuniform vortex matter into an anti-hyperuniform system.
We study density fluctuations in the vortex structure nucleated in the $\beta$-Bi$_{2}$Pd superconductor for large observation length-scales (thousands of vortices) by means of magnetic decoration experiments and Langevin dynamics simulations. Previous experimental studies
at small and intermediate observation lengh-scales  (hundreds of vortices) report that this vortex structure presents gel-like properties directly linked to the disorder potential generated by twist hackle defects.~\cite{Llorens2020b} This type of  correlated disorder is produced in the sample by the propagation of atomic-scale cracks triggered by twisting efforts generated when cleaving the samples.~\cite{Llorens2020b}  Local stresses that alter the superconducting and pinning properties of the sample are induced along the direction of the twist hackles.~\cite{Llorens2020b}
This previous work images the vortex pattern at intermediate densities and reveals the nucleation of rows of vortices pinned at these defects. In some cases, perpendicular rows of vortices are observed. Vortices located at twist hackles act as a scaffold for the nucleation of a structure with voids and clusters and the system presents strong density fluctuations at length-scales of the average vortex spacing, $a_{0} \propto 1/\sqrt{B}$.~\cite{Llorens2020b} This phenomenology typical of a gel-like phase contrasts with the disordered vortex arrangements, but homogeneous at length-scales of $a_{0}$,  observed in media with point-like and correlated columnar disorder.~\cite{Cubitt1993,Blatter1994,Pardo1998,Klein2001,Menghini2002,Fasano2005,Pautrat2007,Petrovic2009,vanderBeek2012,Suderow2014,MarzialiBermudez2015,Zehetmayer2015,ChandraGanguli2015,Toft-Petersen2018,AragonSanchez2019}  Here we study $S(q)$ in the asymptotic $q \rightarrow 0$ limit and reveal that the diluted vortex matter in $\beta$-Bi$_2$Pd is anti-hyperuniform and  argue that this is due to the spatially-correlated density fluctuations and its mode-quenching induced by the particular correlated disorder present in this host media.

\section{Methods}

\subsection{Magnetic decoration imaging}

In order to study  vortex density fluctuations at large lenght-scales we image individual vortex positions in extended fields-of-view (FOV) of thousands of vortices by means of the magnetic decoration technique.~\cite{Fasano2003} We obtain snapshots of the vortex structure by evaporating ferromagnetic particles that are attracted towards the local field gradient entailed by a vortex core.  Figure\,\ref{fig:Figure1} shows a scanning-electron-microscopy image  of a magnetic decoration in $\beta$-Bi$_2$Pd where decorated vortices are observed as black dots. This technique is suitable for discerning individual vortex positions at low densities.~\cite{Fasano2008} This allows us to have
information on the distance-evolution of $\sigma_{N}^{2}$ at large length-scales and also to explore the  small $\mathbf{q}$ limit of $S(\mathbf{q})$ in a range down to two orders of magnitude smaller than the typical wave-vector of the system $q_{0}=2\pi/a_{0}$. We study the case of very diluted vortex densities for magnetic inductions $B=1.8$\,G and 0.9\,G. In the latter case we image up to 17 000 vortices whereas 1 000 vortices are revealed in the former case.

The single crystals of $\beta$-Bi$_{2}$Pd with a critical temperature of 5\,K and a transition width of roughly 0.03\,K were grown from melting with a slight excess
of Bi following the procedure described in Ref.\,~\onlinecite{Herrera2015}. The snapshots of the vortex structure are taken at temperatures of 2.3\,K after performing a field-cooling. During this process, the vortex structure gets frozen at length-scales of $a_{0}$ at a temperature $T_{\rm freez}$ close to the  temperature  at which the  bulk pinning of the samples sets in.~\cite{CejasBolecek2016} From local Hall magnetometry data  we estimate a freezing temperature $T_{\rm freez} \sim 0.97 T_{\rm c}$ for vortex densities of $1-2$\,G. When further cooling down to the decoration temperature, vortices profit from quenched disorder performing excursions on scales of the order of coherence length, one order of magnitude smaller than the spatial resolution of the magnetic decoration of the order of penetration depth $\lambda(0) \sim 0.3 \,\mu$m.~\cite{Soda2022} Then the structure imaged at 2.3\,K corresponds to that frozen when vortex-vortex interactions, depending on the vortex-vortex separation $\sim a_{0}$ normalized by $\lambda(T_{\rm freez}) \sim 0.8$\,$\mu$m. Thus for the studied low densities $a_{0}/\lambda(T_{\rm freez}) = 6$  for 0.9\,G  and 4.2 for 1.8\,G.

\subsection{Langevin dynamics simulations}

\begin{figure}
    \centering
    \includegraphics[width=\columnwidth]{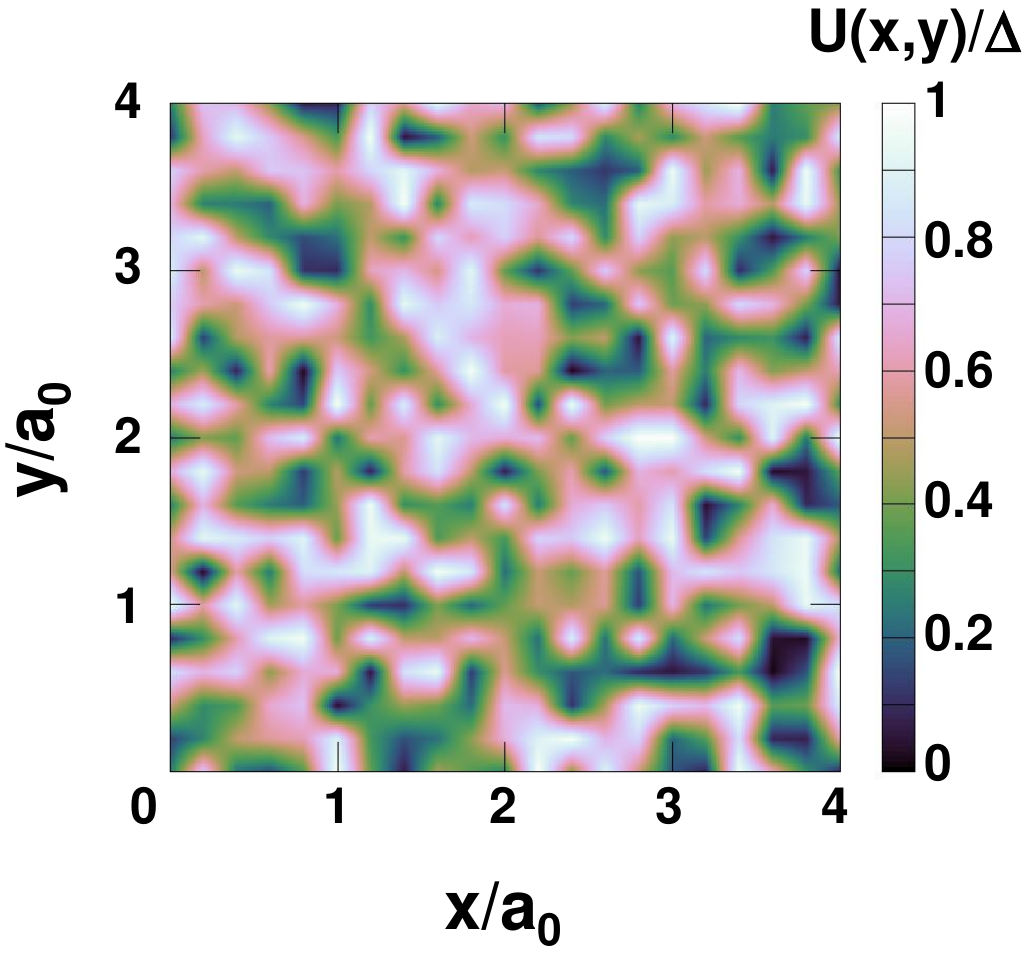}
    \caption{
        Snapshot of a realization of the  short-ranged fourfold correlated disordered potential  $U(x,y)$ described by Eq.(\ref{eq:bilinear}) and used in the vortex dynamics numerical simulations. The image corresponds to the particular choice $l_p=\lambda=0.2 a_{0}$. Colors indicate the magnitude of the potential in units of its maximum amplitude $\Delta$,  $U(x,y)/\Delta$.}

    \label{fig:enter-label}
\end{figure}

\begin{figure*}[ttt]
       \includegraphics[width=2\columnwidth]{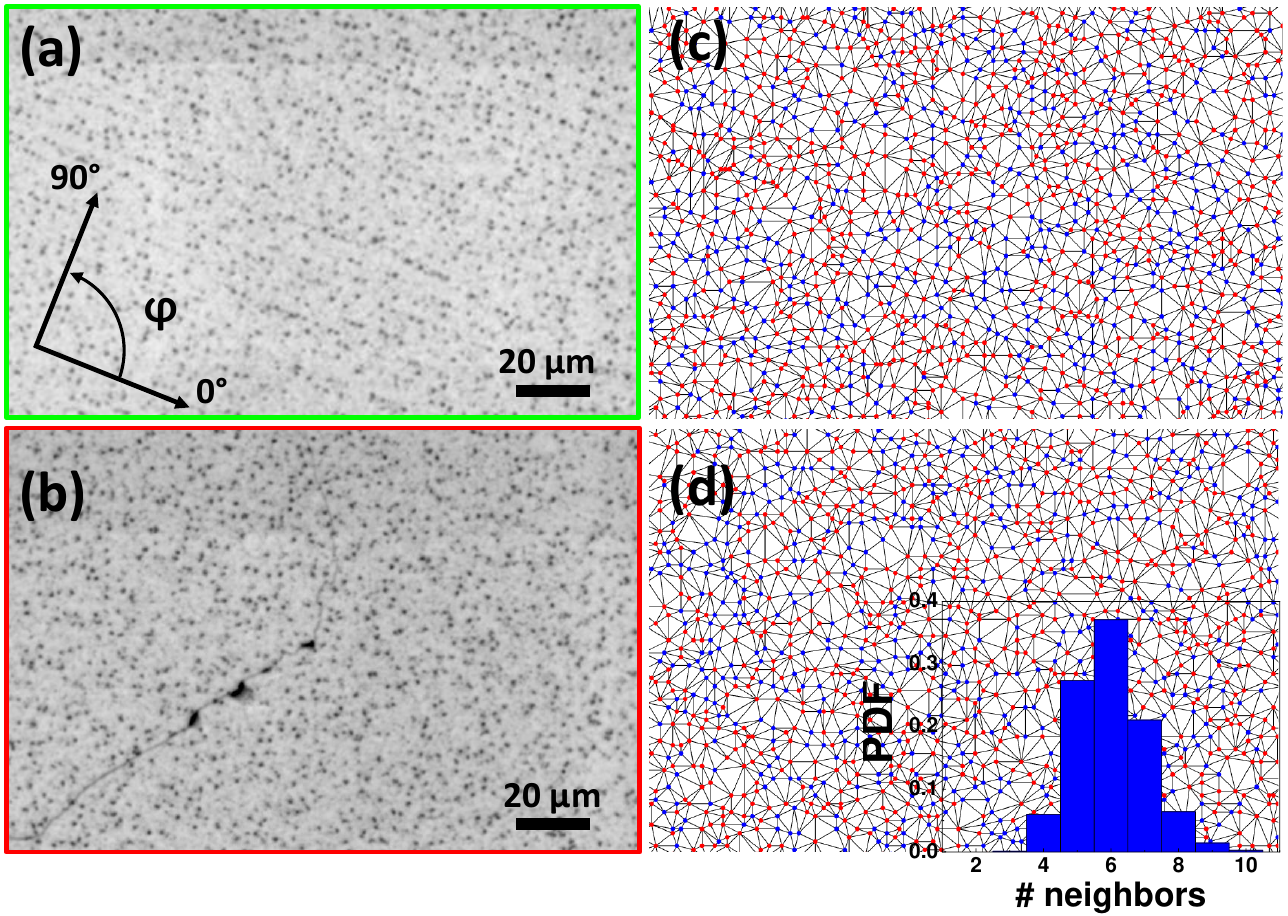}

       \caption{(a-b) Zoom-in of a fraction of the green and red areas highlighted in the panoramic view of the vortex structure nucleated in $\beta$-Bi$_2$Pd at 0.9\,G of Fig.\,\ref{fig:Figure1}. The green region presents rows of vortices aligned along the $\varphi =0$\,$^{\circ}$ and 90$^{\circ}$  directions.  These rows are less evident in the red highlighted region of the sample. (c-d) Delaunay triangulations  of the vortex structures shown in the corresponding left panels
        with neighbor vortices connected by blue lines and non-sixfold coordinated vortices highlighted in red. In both images, as in the rest of the sample, the average density of topological defects is roughly 62\,$\%$. The insert show the histogram of the number of first-neighbors per vortex for the red and green area.}
       \label{fig:Figure2}
\end{figure*}

With the aim of further understanding the vortex density fluctuations observed experimentally in snapshots taken at 2.3\,K, we model interacting vortices in three dimensions.
Since we are interested in the diluted case and the disorder in the samples is expected to be highly correlated along the applied field direction ($z$), we expect most of vortices to be pinned as straight lines aligned with the applied magnetic field. Thus, by symmetry, we can reduce the three dimensional problem to an effective two-dimensional one, ignoring the $z$-coordinate of vortices. We hence performed two-dimensional Langevin dynamics simulations of thousands of interacting vortices with a vortex density comparable to the experimental values. In order to simulate the main characteristics of the vortex structures observed in $\beta$-Bi$_2$Pd, we consider a pinning potential that is short-ranged correlated in the plane and presents fourfold symmetry. The reasons why we considered this symmetry will be evident after details of the observed vortex arrangement discussed at the beginning or the next section.

Specifically, we simulate $N=8192$ interacting vortices in a two-dimensional $L \times L$  pinning landscape with periodic boundary conditions at a finite temperature $T$ and  vortex density $L^2/N \approx 36 \lambda^2$, corresponding to $a_{0}/\lambda \sim 5$. With the aim of emulating the field-cooling experimental protocol, we perform temperature quenches from a highly-disordered vortex structure at  high temperatures towards a lower temperature well below the irreversibility line
where the dynamics becomes nearly frozen. This involves employing a constant and gradual negative-rate temperature-ramp $T=T_0 - b t$, with a small cooling rate $b$, thus allowing the thermal equilibration of the system  up to length-scales of the order of $a_0$.
During this ramp we observe the temporal evolution of the instantaneous structure factor $S(q,t)$ and analyze the over-damped dynamics of $N$ vortices by solving the system of equations
\begin{align}
\eta \frac{d{\bf R}_n}{dt}
 =
&-\frac{\partial}{\partial {\bf R}_n}
\left[
\frac{1}{2}\sum_{m \neq n}^{N} V(|{\bf R}_n-{\bf R}_m|) + U({\bf R}_n)
\right] \nonumber \\
&+\zeta({\bf R}_n,t).
\label{eq:eqmotionsimu}
\end{align}
Here ${\bf R}_n\equiv (X_n, Y_n)$ denotes the positions of vortices $n=1,\dots,N$
and $\eta$ is a viscosity coefficient,
expected to be of the order of the Bardeen-Stephen coefficient.
The interaction potential corresponds to the one for straight vortex lines in a three dimensional medium,
$V({\bf R}) = \epsilon_0 K_0(|{\bf R}|/\lambda)$, with $K_0$ the modified Bessel function of order zero. The Langevin noise satisfies $\langle \zeta({\bf R},t) \rangle=0$ and $\langle \zeta({\bf R},t)\zeta({\bf R}',t') \rangle=2 \eta T \delta({\bf R}-{\bf R}')\delta(t-t')$. The pinning landscape $U({\bf R})$ is a continuous correlated random potential in the plane, obtained by bilinear interpolation in a square grid
with nodes at $(x_n,y_m)\equiv (n,m)l_p$, with $n$ and $m$ integers and $l_p$ a typical distance. At a position ${\bf R}=(x,y)$ such that $x \in [x_n,x_{n+1}]$ and
$y \in [y_m,y_{m+1}]$, the potential reads

\begin{eqnarray}
U(x,y)&=& U_{n,m}(x_{n+1}-x)(x_{m+1}-y) \nonumber\\
&+& U_{n+1,m}(x-x_n)(y_{m+1}-y) \nonumber\\
&+& U_{n,m+1}(y-y_m)(x_{n+1}-x) \nonumber\\
&+& U_{n+1,m+1}(y-y_m)(x-x_n),
\label{eq:bilinear}
\end{eqnarray}
where $U_{n,m}$ are quenched independent random values, $\langle U_{n,m} \rangle=0$ and $\langle U_{n,m}U_{n',m'} \rangle=\Delta^2 \delta_{n,n'}\delta_{m,m'}$, drawn from a uniform distribution.
This two-dimensional potential has an amplitude $\Delta$ (measured in units of $\epsilon_{0}$),  is anisotropically oriented along the $X-Y$ axis, and has a finite correlation length $\propto l_p$. We use $\Delta$ and $l_p$ to tune the effect of this disorder potential in order to reproduce qualitatively the experimental observations on the vortex arrangement in $\beta$-Bi$_2$Pd samples. Figure\,\ref{fig:enter-label} shows a snapshot of a realization of $U(x,y)$ with
the parameter $l_p=\lambda=0.2 a_{0}$ used in the simulations.

Numerical simulation parameters and results are expressed in a dimensionless form. We use $\lambda$ as the unit of length. All energies are expressed in units of $\epsilon_0 d$, and temperature is in units of $\epsilon_0 d/k_B$, where $d$ represents the sample thickness and $k_B$ denotes the Boltzmann constant.
Times are expressed in units of
$\eta \lambda^2/\epsilon_0$.

\section{Results}

\subsection{Experimental results}

Figure\,\ref{fig:Figure1} shows a panoramic image of the very diluted  vortex structure nucleated in $\beta$-Bi$_{2}$Pd samples at 0.9\,G. The field-of-view spans 17 000 vortices imaged as black spots.  At first sight, the structure is rather disordered. Moreover, the vortex arrangement presents voids and clusters of larger density than the average.  In addition, in some regions of the sample, as for instance the green framed one, vortices arrange in perpendicular rows. These rows result from the pinning of vortices in the
twist hackle defects generated when cleaving the sample.~\cite{Llorens2020b}  As observed in the images of Fig.\,\ref{fig:Figure2}\,(a) corresponding to parts of the green and red FOV, some parts of the green area present densely distributed vortex rows alligned in the orthogonal directions $\varphi=0$\,$^{\circ}$ and 90\,$^{\circ}$. In other regions of the sample, such as the red framed one of Fig.\,\ref{fig:Figure2}\,(b), rows of vortices are less evident, suggesting correlated defects are distributed more scarcely in these regions.  Similar results are found for the structure nucleated at 1.8\,G in another sample.

\begin{figure}[hhh]
       \centering
       \includegraphics[width=0.87\columnwidth]{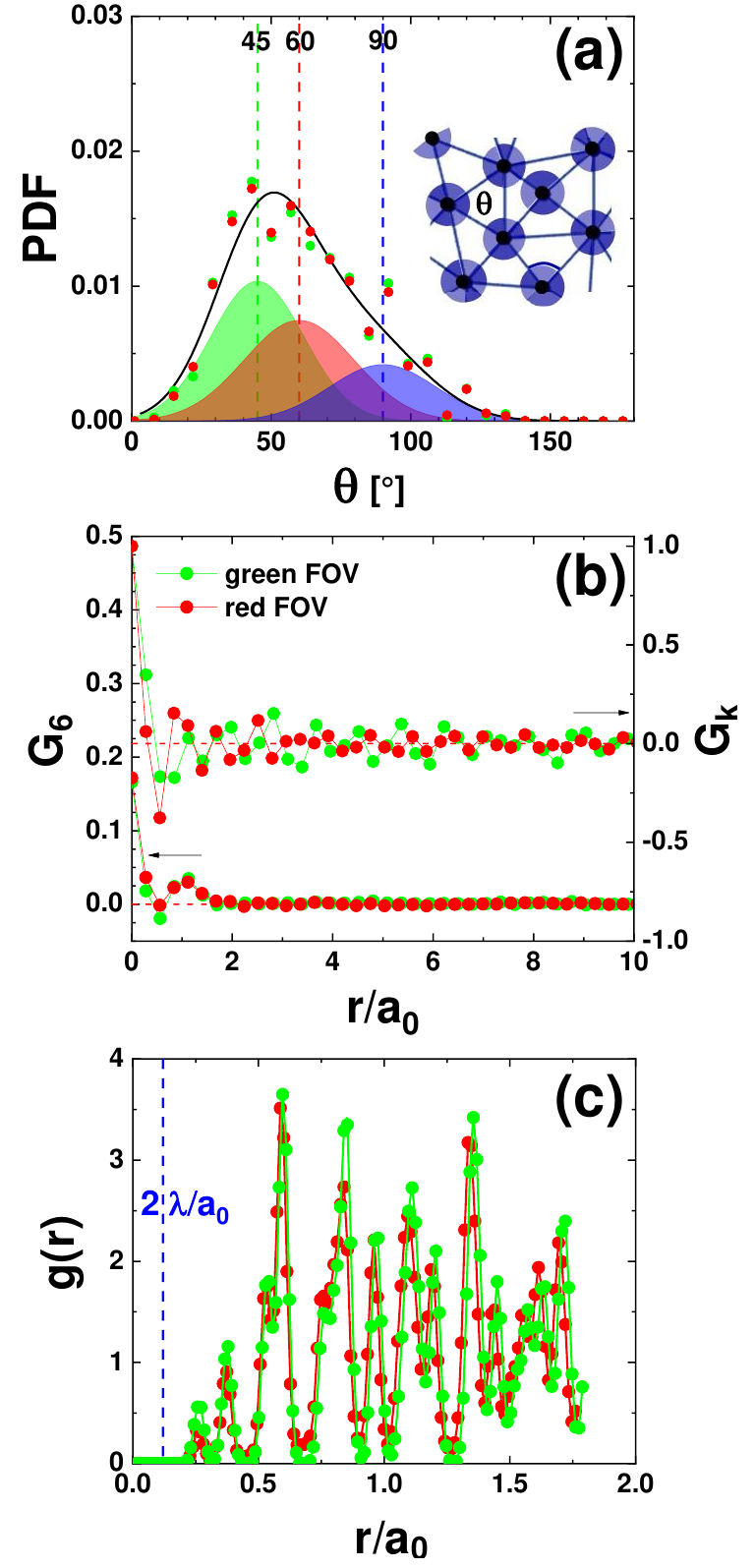}
\caption{(a) Probability density function (PDF) of the internal angles $\theta$ of the triangles of the Delaunay triangulation algorithm that identifies the first neighbors of each vortex, see schematics in the insert. (b) Orientational $G_{\rm 6}$ (bottom, left axis) and positional $G_{\rm K}$ (top, right axis) correlation functions of the vortex structures nucleated in the red and green fields-of-view (FOV) of Fig.\,\ref{fig:Figure1}. Data is shown as function of the distance normalized by the average spacing between first-neighbors, $r/a_{0}$. (c) Pair correlation function $g(r)$ as a function of $r/a_{0}$  for vortices in the red and green FOV. The blue dashed line indicates the typical size of a vortex detected by the magnetic decoration technique, $\sim 2 \cdot\lambda(2.3\,\text{K})$.}
       \label{fig:Figure3a}
\end{figure}

The highly disordered nature of the structure is more evident from the Delaunay triangulation analysis that allows the determination of the amount of sixfold-coordinated vortices as well as topological defects. Figures\,\ref{fig:Figure2} (c) and (d) show the triangulation for the zoomed-in images in green and red regions, respectively. Neighbors are connected by blue lines and non-sixfold coordinated vortices are highlighted in red. In both green and red regions, as well as in the rest of the sample and for the field of 1.8\,G, the average density of non-sixfold coordinated vortices is $\sim 62\,\%$. From the triangles of the Delaunay triangulations we compute the probability density function (PDF) of the distribution of internal angles of the triangles, see Fig.\,\ref{fig:Figure3a} (a). This density is quite similar in the green and red areas with respectively many and scarce correlated defects, see green and red symbols. The distribution of internal angles is well fitted by three Gaussian functions  centered at angles of 45\,$^{\circ}$, 60\,$^{\circ}$ and 90\,$^{\circ}$. The important height of the Gaussians at 45\,$^{\circ}$ and 90\,$^{\circ}$ indicates that a significant fraction of vortices presents local square symmetry. These vortices with four-fold symmetry are typically observed within the rows induced by the correlated disorder present in the samples.

 Density fluctuations in the diluted vortex arrangement nucleated in $\beta$-Bi$_{2}$Pd   are quite important at short length-scales, irrespective of the effect of correlated disorder being strongly or weakly evident in the structure. The standard deviation of the distance between first neighbors is rather large, of around $0.4a_{0}$ for regions green and red, and for both applied fields.  This value is similar to the one measured in $\beta$-Bi$_{2}$Pd at a density of 12.5\,G for 100 vortices applying the SQUID-on-tip (SOT) technique,~\cite{Llorens2020b} and much larger than values found at larger fields in other superconducting materials of $\sim 0.2a_{0}$.~\cite{Guillamon2014}  The important magnitude of vortex density fluctuations at short length-scales can also be deduced from the behavior with distance of the orientational, $G_{\rm 6}$, and positional, $G_{\rm K}$, correlation functions~\cite{Fasano2005} of the structure. Figure\,\ref{fig:Figure3a} (b) shows that $G_{\rm 6}$ and $G_{\rm K}$ stop fluctuating and saturate at zero for $r/a_{0} >2$, an indication that density fluctuations are able to suppress the orientational and positional order at short length-scales.

 \begin{figure}[ttt]
       \centering
       \includegraphics[width=\columnwidth]{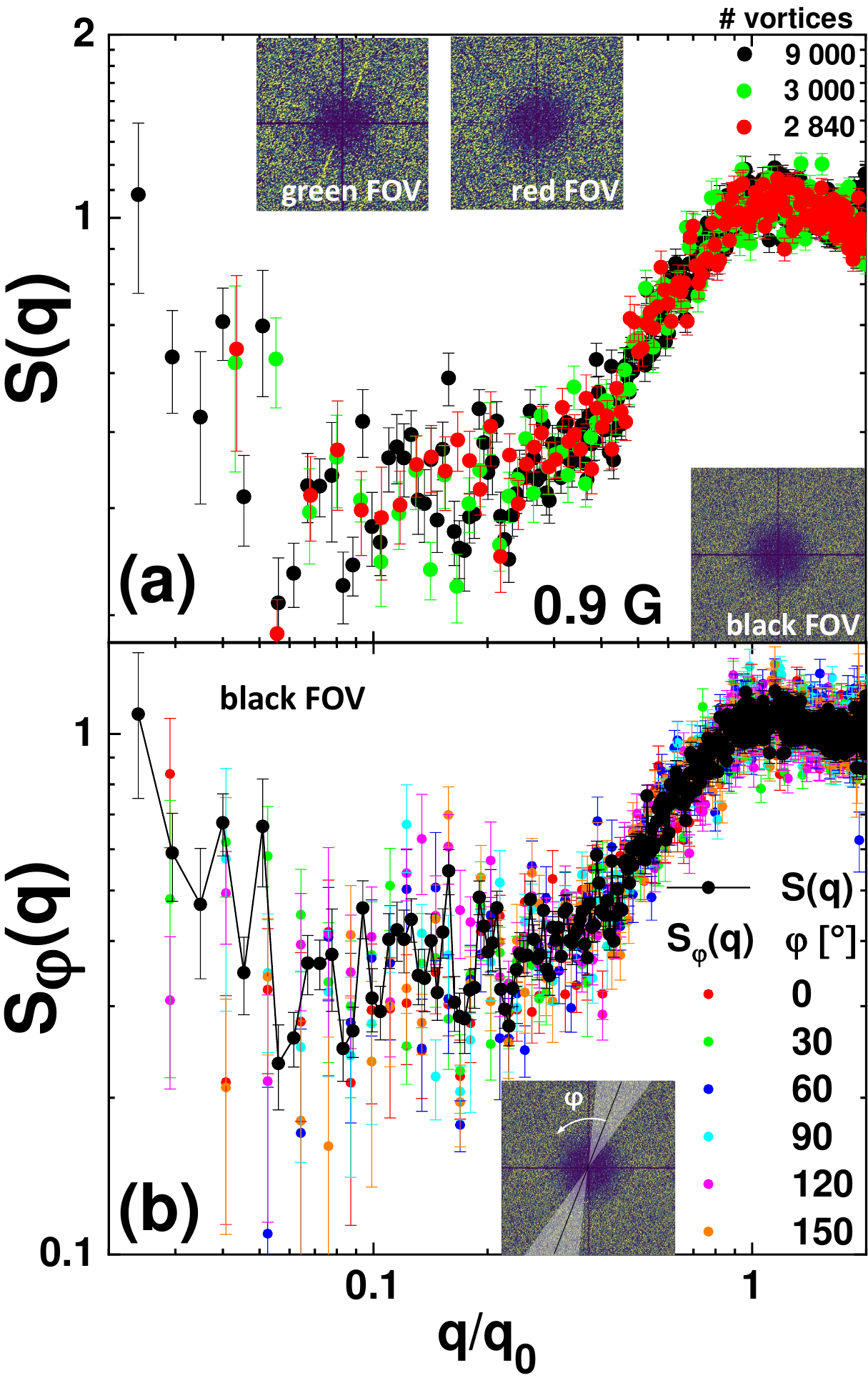}
       \caption{Structure factor data  for the vortex structure nucleated in $\beta$-Bi$_2$Pd at 0.9\,G. (a) Angularly-averaged structure factor $S(q)$ for vortices in the green, red and black field-of-view (FOV) of Fig.\,\ref{fig:Figure1}. Data presented as a function of the wave-vector $q=2\pi/r$ normalized by the wave-vector $q_{0}=2\pi/a_{0}$ with $a_{0}$ the average spacing between first-neighbor vortices.
       Inserts: Two-dimensional structure factors from which the $S(q)$ are computed. The top-right legend   indicates the number of vortices imaged in each FOV. (b) Structure factor data for the black FOV with 9 000 vortices. Color data: Partially-averaged structure factor, $S_{\varphi}(q)$, computed on arcs centered around the $\varphi$ angles indicated in the legend and spanning $\pm 15$\,$^{\circ}$ around these values. Black data: Fully angularly-averaged $S(q)$ data also shown in panel (a).}
       \label{fig:Figure3}
\end{figure}

 A more quantitative analysis of the short-range density fluctuations can be performed by calculating the pair correlation function $g(r)$ that measures the probability of finding a pair of vortices at a distance $r$. In a system with a small amount of density fluctuations $g(r)$ presents sharp maxima at the first, second and so on neighbor distances.  Figure\,\ref{fig:Figure3a} (c) shows the variation of the pair correlation function with $r/a_{0}$ for an extremely diluted structure of vortices with a density of 0.9\,G. The $g(r)$ curves are alike in red and green regions of the sample, presenting four multiple-peaked maxima for $r<a_{0}$. No signal is observed for distances smaller than the typical size of a vortex detected with the magnetic decoration technique, $\sim 2\lambda(2.3\,$K$)$. Multiple peaks of similar height are also observed between $a_{0}$ and  $2a_{0}$. This behavior contrasts  with the typical behavior of a disordered liquid presenting the tallest peak at  $a_{0}$ and then peaks decreasing in intensity for $k \cdot a_{0}$ with $k$ an integer typically between 2 and up to 5-6. Therefore, this behavior of $g(r)$ suggests the system has a tendency towards clustering with  organized density fluctuations at short length-scales.

This disordered structure presenting clustering at typical distances of fractions of $a_{0}$ might be also accompanied, or not, by a hyperuniform hidden order at large length-scales, depending on the degree of spatial correlation of density fluctuations at large length-scales. In order to study this, we calculate the two-dimensional structure factor of the vortex structure nucleated at the surface of the samples, $S(\mathbf{q})=  |\hat{\rho}(q_{\rm x},q_{\rm y},z=0)| ^{2}$.  We compute the $S(\mathbf{q})$ for the green and red regions, as well as for the black region shown in Fig.\,\ref{fig:Figure1} that includes part of the red region but spans in a larger field-of-view. Considering 9 000 vortex positions in the black region allow us to compute $S(\mathbf{q})$ down to smaller wave-vectors.
The $S(\mathbf{q})$ data for the lower density of 0.9\,G shown in the insert to Fig.\,\ref{fig:Figure3} (a) are featureless,  not presenting spots nor a ring of higher intensity. Nevertheless,  the $S(\mathbf{q})$ data in the green FOV shows perpendicular yellow lines of modest intensity due to the arrangement of vortices in perpendicular rows.  For all the studied FOVs, the angularly-averaged structure factor $S(q)$  presents a faint and very broad peak at around the Bragg wave-vector $q_{0}= 2\pi/a_{0}$, see main panel of Fig.\,\ref{fig:Figure3} (a).  This diffuse peak extended in the vicinity of $q_{0}$ is another manifestation of vortex density fluctuations being important at short length-scales.

\begin{figure}[ttt]
       \centering
       \vspace{0.08cm}\includegraphics[width=0.95\columnwidth]{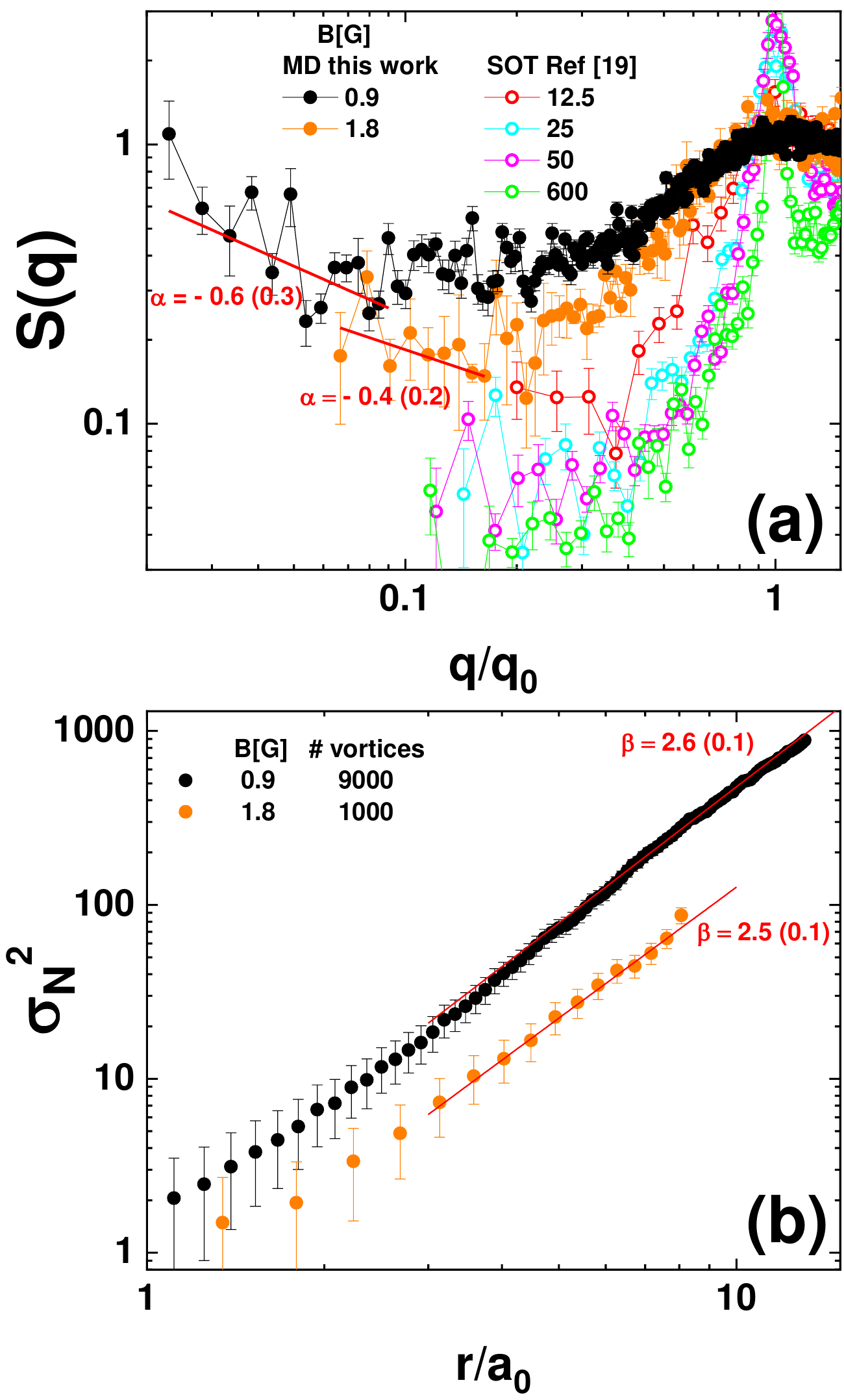}
       \caption{(a) Angularly-averaged structure factor $S(q)$ for vortices nucleated in $\beta$-Bi$_2$Pd at different densities: Magnetic decoration (MD) data from this work at  0.9 and 1.8\,G and SQUID-on-tip (SOT) data for fields up to 600\,G from Ref.\,\onlinecite{Llorens2020b}. The larger fields-of-view in MD data allows us to reveal the asymptotic $S(q)$ behavior down to a $q$-range one order of magnitude smaller than for SOT experiments. Fitting the MD data with an algebraic behavior (red lines) reveals that $S(q)$ grows when decreasing $q$ in the $q \to 0$ limit with exponents $\alpha < 0$ indicated. (b) Variance of the vortex number, $\sigma_{N}^{2}$, as a function of normalized distance $r/a_{0}$ for the MD data considered in (a). Algebraic fits to $\sigma_{N}^{2}$ in the long-wavelength limit (red lines) yield exponents $\beta = 2.6 (0.1)$ for the 0.9\,G structure and   $\beta = 2.5 (0.1)$ for the 1.8\,G structure.}
       \label{fig:Figure4}
\end{figure}

\begin{figure*}[ttt]
       \centering
       \vspace{0.7cm}
\includegraphics[width=1.4\columnwidth]{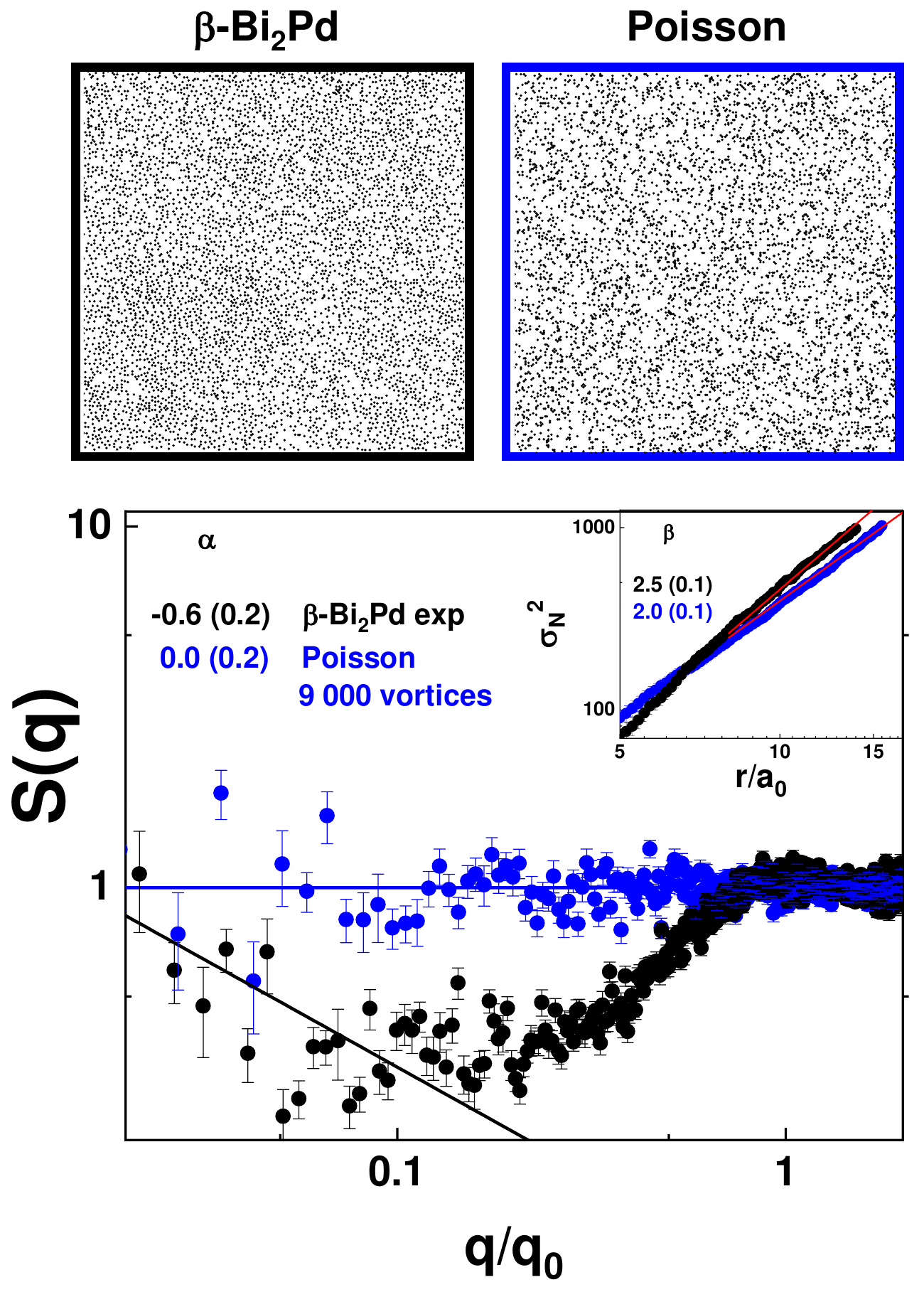}
       \caption{Angularly-averaged structure factor $S(q)$ for the  vortex structure nucleated in $\beta$-Bi$_2$Pd at 0.9\,G (black symbols) and for a simulated  random Poisson distribution of 9 000 points (blue symbols). Top panels: Zoom-in images of the  experimental and simulated structures. Full lines are fits to the $S(q)$ data with algebraic functions in the low-$q$ limit: The experimental data presents an algebraic behavior with exponent $\alpha= -0.6 (0.2)$ whereas the simulated Poisson structure has constant $S(q)$ within the error. Insert: Growing  of the  vortex variance number $\sigma^{2}_{N}$ with normalized distance $r/a_{0}$ for the experimental and simulated structures. Fits performed in the large wavelength limit indicate that for the vortex structure nucleated in $\beta$-Bi$_2$Pd the variance grows with an exponent $\beta=2.5 (0.1)$ larger than for the simulated Poisson structure.}
       \label{fig:Figure5}
\end{figure*}

Regarding the nature of density fluctuations at large length-scales, the data of Fig.\,\ref{fig:Figure3} shows that
for $q/q_{0} < 0.4$  $S(q)$ stops decreasing on lowering $q$ and shows a tendency to stagnation around a pretty large value of 0.6. This behavior is observed in the red, green and black FOVs. This saturation behavior for intermediate wave-vectors is also observed for vortex structures nucleated at larger densities up to 600\,G: Fig.\,\ref{fig:Figure4} (a) shows the $S(q)$ for the structures imaged with the SOT technique in Ref.\,\onlinecite{Llorens2020}. These SOT images contain only up to hundreds vortices and then the small $q$ limit is not accessed in these measurements. In contrast, our magnetic decoration data in the larger black region spans up to 9 000 vortices and allow us to explore the asymptotic $q \rightarrow 0$ limit. The  magnetic decoration data obtained in an extended FOV reveal that, strikingly, in the $q/q_{0} < 0.1$ range the $S(q)$ no longer stagnates but grows on decreasing $q$ and reaches values close to 1 for $q/q_{0} < 0.03$. A similar behavior is also  found in the structure factor data of the vortex structure with a density of 1.8\,G, see orange points in Fig.\,\ref{fig:Figure4} (a). Performing algebraic fits of $S(q)$ in the short wave-vector range yields exponents $\alpha= - 0.6 (0.3)$ and $- 0.4 (0.2)$ for the diluted densities of 0.9 and 1.8 G, respectively.

This phenomenology is a fingerprint of anti-hyperuniform systems and indicates that at large length-scales the density fluctuations grow faster than
for a non-hyperuniform fluid.
Indeed, Fig.\,\ref{fig:Figure4} (b) shows that the standard deviation of the number of vortices  grows algebraically with the size of the observation window in the large length-scale limit
with exponents $\beta=2.6(0.1)$ and $2.5(0.1)$ for densities of 0.9 and 1.8\,G, respectively. The agreement between the fitted exponents for $S(q)$ and $\sigma_{\rm N}^{2}(r)$ is quantitatively good since a relation $\alpha = 2 - \beta$ is expected.

The increase of $S(q)$ when decreasing $q$
in the $q \rightarrow 0$ limit is isotropic in the azimuthal angle. This can be deduced from the $S_{\varphi}(q)$ data of
Fig.\,\ref{fig:Figure3} (b) for vortices in the black FOV that shows structure factor data partially-averaged in angle  in arcs of $\pm 15\,^{\circ}$ around the  $\varphi$ values indicated in the legend. All the $S_{\varphi}(q)$ curves, irrespective of being along or outside of the directions of the planar defects ($\varphi=0$ and $90\,^{\circ}$), show a tendency to increase when decreasing $q/q_{0}$ below 0.1. Thus, even though the disorder has a preferential orientation of perpendicular planes, the density fluctuations of the elastic system nucleated in such a host media isotropically grow when increasing the length-scale in an anti-hyperuniform fashion.

\begin{figure}
    \centering
    \includegraphics[width=0.95\columnwidth]{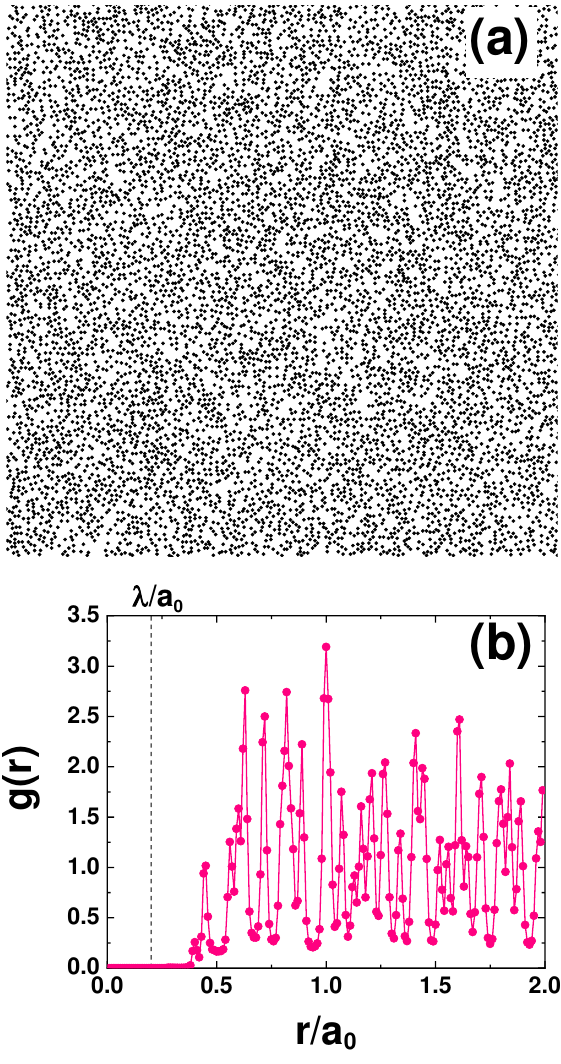}
    \caption{Langevin dynamics simulation results for $8192$ interacting vortices with a diluted density of $\sim 1$\,G, $a_{0}/\lambda = 5$, nucleated in a sample with  the fourfold-symmetric and short-ranged correlated disorder potential  following a field-cooling protocol similar to the experimental one. (a) Typical snapshot of one of the quenched configurations after a cooling ramp time $t=10 000$ obtained in one realization of the correlated disorder. (b) Pair distribution function $g(r)$ of the quenched configuration averaged over $100$ different realizations of the correlated disorder.}
    \label{fig:gdrsim}
    \end{figure}

The anti-hyperuniform diluted vortex structure nucleated in $\beta$-Bi$_2$Pd
presents  density fluctuations at large length-scales that grow even faster than those expected  for elastic objects distributed at random. Indeed, Fig.\,\ref{fig:Figure5} shows the comparison between  $S(q)$ and $\sigma_{\rm N}^2$ for the experimental data (black symbols) and a simulated Poisson distribution of 9 000 points (blue symbols). The spatial arrangement of objects seems alike in both cases, see top panels of   Fig.\,\ref{fig:Figure5}, and clustering is observed in both structures. Moreover, the magnitude of density fluctuations at short and intermediate length-scales is larger for the simulated random than the experimental structure, see the $\sigma_{\rm N}^2$ blue curve located above the black one for the $r/a_{0} < 7$ range in the insert to Fig.\,\ref{fig:Figure5}. This is due to the short-range repulsion between vortices that allows for clustering but respecting a minimal safe separation between them.

However, at large length-scales
 the experimental structure presents a more rapid growth of
 the number variance
 than the random case: $\sigma_{\rm N}^2$ grows as $r^{2}$ in the random case whereas $\beta$ is larger than two. Accordingly, the $S(q)$ of the random structure fluctuates around one (since it is a single realization over disorder) whereas for the experimental vortex structure grows on decreasing $q$ in the asymptotic $q \to 0$ limit.
 Thus the nucleation of this anti-hyperuniform vortex matter is mainly due to spatially-correlated density fluctuations induced at large length-scales by the peculiar disorder of the host media. We would like to recall here that all these findings are obtained from snapshots resulting after a field-cooling of the vortex structure. In what follows, we will discuss how  mode-freezing effects also contribute to enhance the anti-hyperuniformity of the system by means of simulations of the experimental field-cooling protocol.

\subsection{Simulation Results}

With the aim of understanding how the anti-hyperuniform properties of the snapshots of the vortex structure nucleated in $\beta$-Bi$_2$Pd emerge during the experimental cooling protocol,
we performed Langevin dynamics simulations of interacting vortices nucleated in a media with a disorder potential that mimics the one present in real samples. As described in Methods, we simulate the temperature quench of the vortex structure starting from a temperature well above the freezing temperature of vortices. Figure\,\ref{fig:gdrsim}\,(a) shows a zoom-in of a snapshot of a typical final configuration while panel (b) presents
the pair correlation function $g(r)$ averaged over 100 final configurations obtained each with a different disorder realization. Simulated snapshots present voids and oriented clusters as similarly observed in the experimental data. Moreover, $g(r)$ presents several peaks for distances below $a_0$, signaling clustering, in qualitative agreement with the experimental data of  Fig.\ref{fig:Figure3a}(c).

\begin{figure}[]
\centering
\includegraphics[width=0.98\columnwidth]{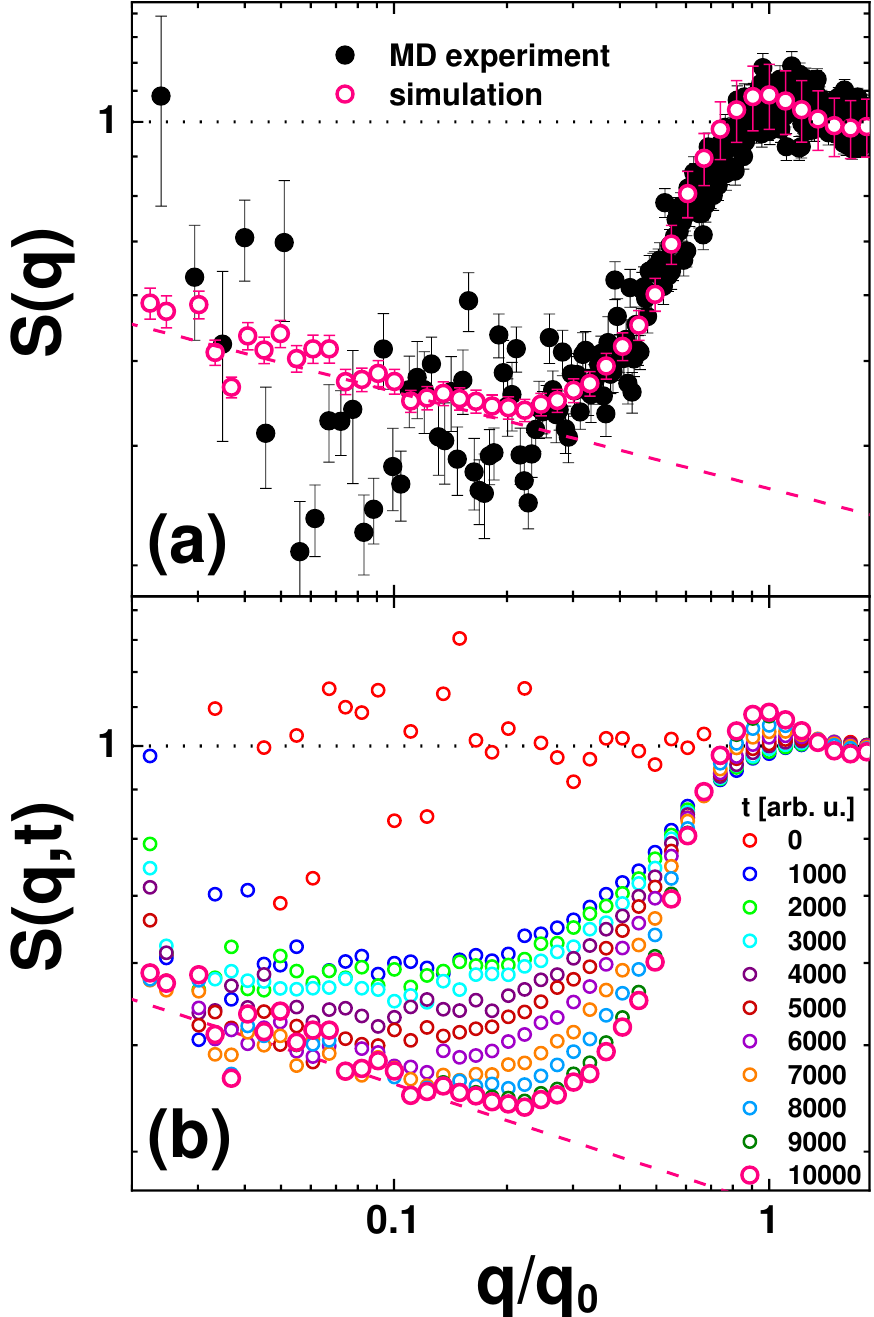}
       \caption{(a) Angularly-averaged structure factor data, $S(q)$, for the experimental vortex structure nucleated at $B=0.9$\,G (black symbols) and for the final configuration obtained  from simulations at roughly the same vortex density and a cooling time $t= 10000$ (open magenta symbols). The dashed line is a power-law fit to the data obtained in simulations in the low-$q$ regime yielding a exponent $\alpha = -0.14 (0.03)$. (b) Cooling-time dependent structure factor data $S(q,t)$ obtained from simulations during the temperature quench. The initial condition corresponds to high temperatures and the dotted-line shows the behavior expected for a random Poisson distribution of vortices. Mode-relaxation gets practically frozen in a metastable configuration at long times, also corresponding to low temperatures. The legend indicates the cooling times in arbitrary units of $\eta \lambda^2/\epsilon_0$, and the largest time corresponds to the curve shown in (a) used to compare with experimental data. Every single one of the curves is the average of $S(q,t)$ over 100 different realizations of the correlated disorder potential.
       }
\label{fig:sim1}
\end{figure}

Figure \ref{fig:sim1}(a) shows the quantitative agreement between the structure factor data in experiments and simulations after tuning the pinning strength {$\Delta = 0.7\,\epsilon_{0}$} and correlation length {$l_p = \lambda$} in order to closely reproduce the experimental arrangement.  Figure \ref{fig:sim1}(b) shows the time-dependent structure factor obtained in the simulations, $S(q,t)$. Initially, $S(q,t=0)\approx 1$, reflecting the nearly-Poissonian spatial distribution of vortices at the high temperatures of the liquid phase at which simulations and experiments start.
As the time increases, and accordingly the temperature decreases, the vortex structure gradually becomes more ordered at the shortest length-scales or equivalently largest $q$. This evolution is evident through the development of a peak around $q_0$. For $q \lesssim q_0$, $S(q,t)$ presents a minimum followed by an ascending effective power-law for $q \to 0$. The minimum at intermediate $q$ values becomes deeper on increasing time. Irrespective of time, in the $q \to 0$ limit, $S(q,t)$ depicts a tendency to grow on decreasing $q$ but for the smallest $q$ values accessed approaches values below but close to $1$.

Figure\,\ref{fig:sim1}(a) shows that the structure factor in our simulations reasonably follow the experimental data for the longest simulated time of
$t = 10 000 \,\eta \lambda^2/\epsilon_0$.
We performed simulations for different system sizes and we systematically observe $S(q,t)<1$ at low $q$ in the
$0.02 \lesssim q/q_0 \lesssim 0.15$ range. We also find that the effective exponent $\alpha= - 0.14$ fitted from an average over 100 disorder realizations of $S(q,t)$ is actually not universal but appears to vary with the pinning strength and time. Indeed, in the extreme case of absence of disorder the effective exponent is much larger, see  the Appendix for details.
These subtle dependencies with disorder strength and time may explain the slight discrepancy between the simulation ($\alpha \approx -0.6 \pm 0.2$) and the experimental results in the low $q$ regime. Indeed, it is difficult to match the simulations and experimental time-scales.
Nonetheless, both the experimental and the simulation effective exponents are negative, indicating a significant growth of density fluctuations when decreasing $q$.

Finally, it is worth noting that the temporal evolution shown in Fig.\,\ref{fig:sim1}\,(b),
and in particular the decrease of $S(q,t)$ with increasing $q$ for small $q$ is qualitatively similar to the one expected in rather minimalist models. For instance, if we linearize the relaxation dynamics of the vortex density fluctuations, modes become independent and the relaxation of modes are exactly described by
\begin{align}
S(q,t)= (S(q,0)-S(q,\infty))f(t/\tau_q)
+ S(q,\infty)
\label{eq:qrel}
\end{align}
where $\tau_{q}$ is the characteristic relaxation time for the mode with wave-vector $q$, $f(x)$ is a function such that $f(x=0)=1$ and $f(x\to \infty)\to 0$, and $S(q,\infty)$ is the equilibrium structure factor at very long times.
The relaxation time $\tau_q$ in this model is expected to increase rapidly with decreasing $q$,
typically as $\tau_q \sim q^{-z}$, with a dynamic exponent $z>0$.
In addition, due to mode relaxation, $S(q,\infty)<S(q,0)$ is expected at small $q$ since initial density fluctuations are suppressed.
Then, if we observe the structure at a particular time $t=t_{\rm freez}$
during the cooling ramp at which the freezing temperature is reached, namely $T_{\rm freez} \equiv T(t_{\rm freez})$ as in the experiment, then the system will be able to locally equilibrate only for length-scales $\sim q^{-1}$ such that $\tau_q<t_{\rm freez}$.
If we assume a constant $S(q,\infty)<1$ at low $q$, and $\tau_q \sim q^{-z}$ with $z>0$, the last equation necessarily implies that $S(q,t)$ must increase with decreasing $q$ at low $q$, since $S(q,0)=1$. Therefore, although a dynamical model of independent modes is a naive approximation, both for the experimental system and the  simulated model, it qualitatively illustrates the fact that a non-equilibrium effect alone can explain a growth of $S(q,t)$ with decreasing $q$ during the freezing process.

Although disorder is not essential for explaining the growth of $S(q,t)$ with decreasing $q$ in the low $q$ range,
it is nevertheless relevant since amplifies the time-scales of the relaxation,
affects the precise shape of $S(q,t)$, and explains salient experimental observations such as the clustering evidenced in $g(r)$ (cf. with the case without disorder shown in Fig.\ref{fig:Figure10} of the Appendix).
Finally, it is worth pointing out that a quench dynamics controlled by a single growing correlation length, as described by the naive model behind Eq. \ref{eq:qrel}, has  nevertheless been observed in the more complicated thermally-assisted dynamics of elastic manifolds in the presence of quenched disorder.~\cite{Kolton2005}

\section{Conclusions}

Non-hyperuniform systems in general are abundant in nature since all compressible systems are non-hyperuniform at thermal equilibrium. Interestingly,
vortex matter nucleated in a host media with point-like disorder has the peculiarity that, although the three-dimensional system has finite compressibility and hence bounded large wavelength density fluctuations at equilibrium, the arrangement of vortices in any plane of the sample is hyperuniform.~\cite{Rumi2019}  However, in the same system of interacting objects hyperuniformity is depleted by finite-thickness effects~\cite{Rumi2019,Besana2024} or even anisotropically suppressed if the host media presents strong planar correlated disorder.~\cite{Puig2022} Interestingly, for thick enough samples the out of equilibrium structures obtained by a temperature quench can avoid such depletion by freezing and memorizing the large length-scale hyperuniform order of the high temperature phase, where disorder is typically less important than vortex-vortex interactions~\cite{Rumi2019}.

In view of these previous findings, and of the evidence presented here, what renders the system of diluted vortices nucleated in $\beta$-Bi$_2$Pd samples
distinctive is its  anti-hyperuniform
and  clustering properties.
The peculiar type of correlated disorder in these samples induces clustering that is revealed in the pair correlation function $g(r)$ as several peaks below $a_0$, suggesting an intra-cluster order.
The anti-hyperuniformity of vortex matter in $\beta$-Bi$_2$Pd has its origin in the non-equilibrium nature of the structures observed and
is of a weak type, as density fluctuations remain bounded in the thermodynamic limit, in contrast with fractals where $S(q)$ is expected to diverge with the system size as $q\to 0$.

In summary, by studying density fluctuations in images with a large number of vortices at diluted densities we reveal the system is anti-hyperuniform
due to the influence of short-ranged correlated defects in the out of equilibrium quenched structure.
This result contrasts other kind of influences, like point defects, which instead favor ordered or disordered hyperuniform behaviors. The vortex lattice is a system of particles whose interactions are controlled by the physical properties of the superconductor carrying vortices. Thus, vortex matter is a rich model system, which permits the generation of hyperuniform or non-hyperuniform structures, depending on the type, correlated \textit{vs.} uncorrelated, and strength of the disorder, the density of the interacting constituents, the thickness of the samples, and the quenching dynamics from high temperatures to the observation temperatures.

\section{Appendix:Numerical Simulations without disorder}
\label{sec:numerics}

\begin{figure}
    \centering
\includegraphics[width=0.9\columnwidth]{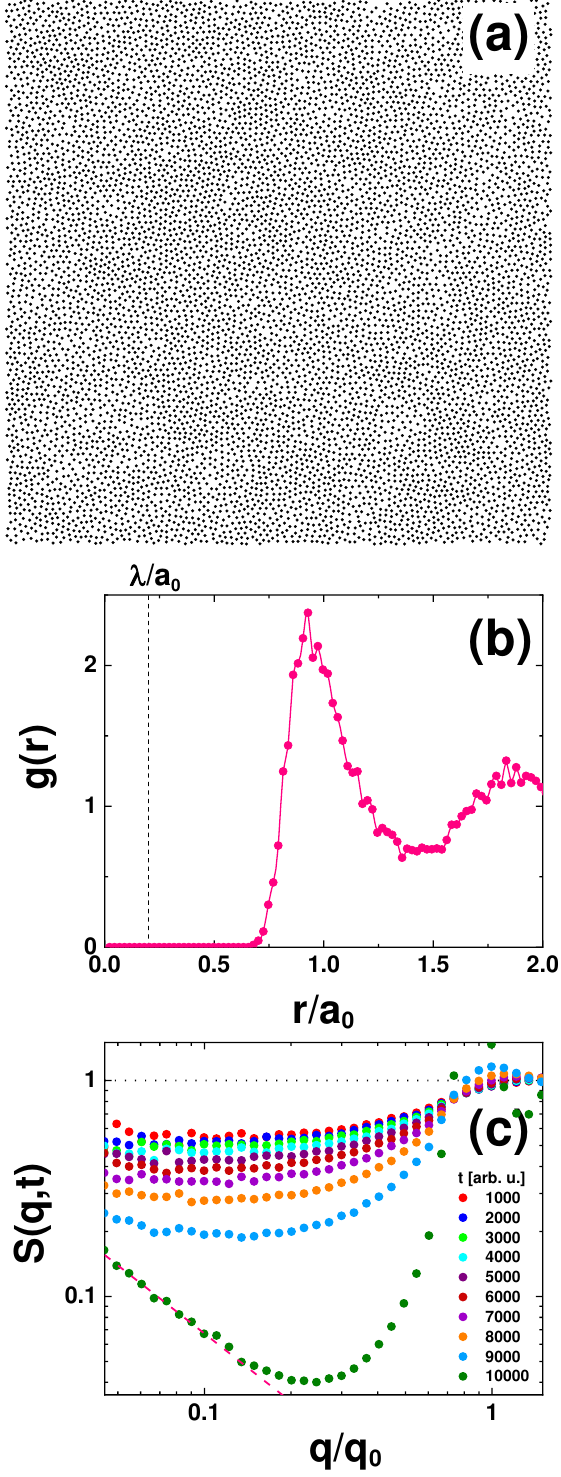}
    \caption{
    Langevin dynamics simulation results for a superconducting sample without disorder. We considered 8192 vortices with a diluted density of $\approx 1$\,G and $a_{0}/\lambda =5$ as in the case of the simulations discussed in the main text.
    (a) Typical snapshot of one of the quenched field-cooling configurations after a cooling ramp time $t=10000$. (b) Pair distribution function $g(r)$ of the snapshot of the top panel.(c) Cooling-time dependent structure factor data $S(q,t)$ obtained at different stages of the temperature quench. Dotted line: behavior expected for a random Poisson distribution of vortices.
     }
\label{fig:Figure10}
\end{figure}

We present here numerical results for the case without disorder by solving Eq. \eqref{eq:eqmotionsimu} with $\Delta=0$ for the same initial conditions, protocol, and vortex density than data in Figures \ref{fig:gdrsim} and \ref{fig:sim1}.
Figure\,\ref{fig:Figure10}\,(a) shows that the final configuration in this case is disordered. Moreover, the $g(r)$ data of Fig.\,~\ref{fig:Figure10}(b) indicates that the structure is liquid-like
and does not present peaks for $r<a_0$, in contrast to what is observed in experiments and simulations with the peculiar type of disorder of Eq.\ref{eq:bilinear}.
The $S(q,t)$, besides evidencing a much more ordered structure for large $q$, growths with decreasing $q$ in the low-$q$ limit with a power-law exponent $\alpha \approx 1$ at $t=10000$ (see pink dashed line in Fig.\,\ref{fig:Figure10}\,(c)). This exponent is significantly larger than the ones observed in experiments  and simulations with disorder.

\section{Acknowledgements}

We thank H. M{\"o}nchshof for insightful discussions. Work supported by the National Council of Scientific and Technical Research of Argentina (CONICET)
through grant PIP 2021-1848, by the  Argentinean Agency for the Promotion of Science and Technology (ANPCyT) through grants PICT 2018-1533 and PICT 2019-1991, and by the Universidad
Nacional de Cuyo research Grants 06/C008-T1
and 06/C014-T1. We also acknowledge support by the Spanish Research State Agency through grants PID-2020-114071RB-I00, PID-2020-117671GB-100, PDC2021-121086-I00, TED2021-130546B-I00 and CEX2018-000805-M, and the EU through grant agreement No. 871106 and COST Action CA21144. Work also supported by the Slovak Research and Development Agency, under Grant No. APVV-20-0425, by the Slovak Scientific Grant Agency under Contract VEGA-2/0073/24 and by the U.S. Steel Kosice, s.r.o.
Y. F. thanks funding from the Alexander von Humboldt Foundation through the Georg Forster Research Award and from the Technische Universit\"{a}t Dresden through the Dresden Senior Fellowship Program.\\


\end{document}